# Positive data kernel density estimation via the logKDE package for R

Andrew T. Jones[1], Hien D. Nguyen[2], and Geoffrey J. McLachlan[1]

[1] University of Queensland, St. Lucia 4072, QLD Australia,
[2] La Trobe University, Bundoora 3086, VIC Australia,
h.nguyen5@latrobe.edu.au.

**Abstract.** Kernel density estimators (KDEs) are ubiquitous tools for nonparametric estimation of probability density functions (PDFs), when data are obtained from unknown data generating processes. The KDEs that are typically available in software packages are defined, and designed, to estimate real-valued data. When applied to positive data, these typical KDEs do not yield *bona fide* PDFs. A log-transformation methodology can be applied to produce a nonparametric estimator that is appropriate and yields proper PDFs over positive supports. We call the KDEs obtained via this transformation log-KDEs. We derive expressions for the pointwise biases, variances, and mean-squared errors of the log-KDEs that are obtained via various kernel functions. Mean integrated squared error (MISE) and asymptotic MISE results are also provided and a plug-in rule for log-KDE bandwidths is derived. We demonstrate the log-KDEs methodology via our $R$ package, `logKDE`. Real data case studies are provided to demonstrate the log-KDE approach.

## 1 Introduction

Let $X$ be a random variable that arises from a distribution that can be characterized by an unknown density function $f_X(x)$. Assume that (A1) $X$ is supported on $\mathbb{R}$, and (A2) $f_X(x)$ is sufficiently continuously differentiable (i.e. $\int_{\mathbb{R}} |f^{(m)}(x)| \, \mathrm{d}x < \infty$, where $f^{(m)}(x)$ is the $m$th derivative of $f(x)$, for $m \leq M \in \mathbb{N}$).

Let $\{X_i\}_{i=1}^n$ be an independent and identically distributed (IID) sample of random variables, where each $X_i$ is identically distributed to $X$ ($i \in [n] = \{1, \ldots, n\}$). Under conditions (A1) and (A2), a common approach to estimating $f_X(x)$ is via the kernel density estimator (KDE)

$$\hat{f}(x) = \frac{1}{nh} \sum_{i=1}^n K\left(\frac{x - X_i}{h}\right), \qquad (1)$$

which is constructed from the sample $\{X_i\}_{i=1}^n$. Here, $K(x)$ is a probability density function on $\mathbb{R}$ and is called the kernel function, and $h > 0$ is a tuning pa-rameter that is referred to as the bandwidth. This approach was first proposed in the seminal works of [12] and [14].

Assume (B1) $\int_{\mathbb{R}} K(x) \, \mathrm{d}x = 1$, (B2) $\int_{\mathbb{R}} xK(x) \, \mathrm{d}x = 0$, (B3) $\int_{\mathbb{R}} x^2 K(x) \, \mathrm{d}x = 1$, and (B4) $\int_{\mathbb{R}} K^2(x) \, \mathrm{d}x < \infty$, regarding the kernel function $K(x)$. Under conditions (A1) and (A2), [12] showed that (B1)–(B4) allowed for useful derivations of expressions for the mean squared error (MSE) and mean integrated squared error (MISE) between $f_X(x)$ and (1); see for example [17, Ch. 3] and [18, Ch. 24]. Furthermore, simple conditions can be derived for ensuring pointwise asymptotic unbiasedness and consistency of (1) (cf. [5, Sec. 32.7]). See [19] for further exposition regarding kernel density estimation (KDE).

The estimation of $f_X(x)$ by (1) has become a ubiquitous part of modern data analysis, data mining, and visualization (see e.g., [1, Sec. 8.6] and [24, Sec. 9.3]). The popularity of the methodology has made its implementation a staple in most data analytic software packages. For example, in the $R$ statistical programming environment [13], KDE can be conducted using the core-package function `density`.

Unfortunately, when (A1) is not satisfied and is instead replaced by (A1*) $X$ is supported on $(0, \infty)$, using a KDE of form (1) that is constructed, from a kernel that satisfies (B1)–(B4), no longer provides a reasonable estimator of $f_X(x)$. That is, if $K(x) > 0$ for all $x \in \mathbb{R}$, and (B1)–(B4) are satisfied, then $\int_0^\infty \hat{f}_X(x) \, \mathrm{d}x < 1$ and thus (1) is no longer a proper *bona fide* (PDF) over $(0, \infty)$. For example, this occurs when $K(x)$ is taken to be the popular Gaussian (normal) kernel function. Furthermore, expressions for MSE and MISE between $f_X(x)$ and (1) are no longer correct under (A1*) and (A2).

In [4], the authors proposed a simple and elegant solution to the problem of estimating $f_X(x)$ under (A1*) and (A2). First, let $Y = \log X$, $Y_i = \log X_i$ ($i \in [n]$), and $f_Y(y)$ be the PDF of $Y$. Note that if $X$ is supported on $(0, \infty)$ then the support of $f_Y(y)$ satisfies (A1). If we wish to estimate $f_Y(y)$, we can utilize a KDE of form (1), constructed from $\{Y_i\}_{i=1}^n$, with a kernel that satisfies (B1)–(B4). If $f_Y(y)$ also satisfies (A2), then we can calculate the MSE and MISE between $f_Y(y)$ and (1).

Let $W$ be a random variable and $U = G(W)$, where $G(w)$ is a strictly increasing function. If the distribution of $U$ and $W$ can be characterized by the PDFs $f_U(u)$ and $f_W(w)$, respectively, then the change-of-variable formula yields: $f_W(w) = f_U(G(w)) G^{(1)}(w)$ (cf. [2, Thm. 3.6.1]). Utilizing the fact that $\mathrm{d} \log x / \mathrm{d}x = x^{-1}$, [4] used the aforementioned formula to derive the log-kernel density estimator (log-KDE)

$$\begin{aligned}
\hat{f}_{\log}(x) &= x^{-1} \hat{f}_Y(\log x) \\
&= \frac{1}{nh} \sum_{i=1}^n x^{-1} K\left(\frac{\log x - \log X_i}{h}\right) \\
&= \frac{1}{n} \sum_{i=1}^n L(x; X_i, h),
\end{aligned} \qquad (2)$$

where $L(x; z, h) = (xh)^{-1} K\left(\log\left[(x/z)^{1/h}\right]\right)$ is the log-kernel function with bandwidth $h > 0$, at location parameter $z$. For any $z \in (0, \infty)$ and $h \in (0, \infty)$, $L(x; z, h)$ has the properties that (C1) $L(x; z, h) \geq 0$ for all $x \in (0, \infty)$ and (C2) $\int_0^\infty L(x; z, h) \, dx = 1$, when (B1)–(B4) are satisfied.

By property (C2), we observe that $\int_0^\infty \hat{f}_X(x) \, dx = 1$, thus making (2) a *bona fide* PDF on $(0, \infty)$. Furthermore, using the expressions for the MSE and MISE between $f_Y(y)$ and (1), we can derive the relevant quantities for (2) as well as demonstrate its asymptotic unbiasedness and consistency.

For every kernel function that satisfies (B1)–(B4), there is a log-kernel function that satisfies (C1) and (C2), which generates a log-KDE that is a proper PDF over $(0, \infty)$. We have compiled an array of other potential pairs of kernel and log-kernel functions in Table 1. Throughout Table 1, the function $\mathbb{I}\{A\}$ takes value 1 if statement $A$ is true and 0, otherwise.

**Table 1.** Pairs of kernel functions $K(y)$ and log-kernel functions $L(x; z, h)$, where $z \in (0, \infty)$ and $h \in (0, \infty)$.

| Kernel | $K(y)$ |
| --- | --- |
| Epanechnikov | $3(5 - y^2) / (20\sqrt{5}) \, \mathbb{I}\{y \in (-\sqrt{5}, \sqrt{5})\}$ |
| Gaussian (normal) | $(2\pi)^{-1/2} \exp(-y^2/2)$ |
| Laplace | $(\sqrt{2}/2) \exp(-2^{1/2} |y|)$ |
| Logistic | $(\pi/4\sqrt{3}) \operatorname{sech}^2(\pi y/2\sqrt{3})$ |
| Triangular | $(\sqrt{6} - |y|) \, 6^{-1} \mathbb{I}\{y \in (-\sqrt{6}, \sqrt{6})\}$ |
| Uniform | $(2\sqrt{3})^{-1} \mathbb{I}\{y \in (-\sqrt{3}, \sqrt{3})\}$ |
| **Log-Kernel** | $L(x; z, h)$ |
| Log-Epanechnikov | $3(5 - x^2) / (20\sqrt{5} x h) \, \mathbb{I}\left\{\log\left[(x/z)^{1/h}\right] \in (-\sqrt{5}, \sqrt{5})\right\}$ |
| Log-Gaussian | $(2\pi)^{-1/2} (xh)^{-1} \exp\left(-\log^2\left[(x/z)^{1/h}\right]/2\right)$ |
| Log-Laplace | $(\sqrt{2}/2)(xh)^{-1} \exp\left(-2^{1/2} \left|\log\left[(x/z)^{1/h}\right]\right|\right)$ |
| Log-Logistic | $(\pi/4\sqrt{3})(xh)^{-1} \operatorname{sech}^2\left(\pi \log\left[(x/z)^{1/h}\right]/2\sqrt{3}\right)$ |
| Log-Triangular | $(xh)^{-1} \left(\sqrt{6}/6 - |x|\right) 6^{-1} \mathbb{I}\left\{\log\left[(x/z)^{1/h}\right] \in (-\sqrt{6}, \sqrt{6})\right\}$ |
| Log-Uniform | $(2\sqrt{3} x h)^{-1} \mathbb{I}\left\{\log\left[(x/z)^{1/h}\right] \in (-\sqrt{3}, \sqrt{3})\right\}$ |

Unfortunately, out of all of the listed function pairs from Table 1, only the normal and log-Gaussian (log-normal) PDFs have been considered for use as kernel and log-kernel functions, respectively, for conducting log-KDE. The log-normal PDF was used explicitly for the construction of log-KDEs in [4], and more generally, for conducting asymmetric KDE on the support $(0, \infty)$, in [9]. Other works that have considered the log-normal PDF for conducting asymmetric KDE include [6], [8], [7], and [22]. In the $R$ environment, asymmetric KDE with lognormal PDF as kernels has been implemented through the `dke.fun` function from the package `Ake` [22].

In this paper we firstly expand upon the theoretical results that were reported in [4], who derived expressions for the biases and variances between generic log-KDEs and their estimands. Here, we utilize general results for KDE of transformed data from [10] and [20]. We further derive a plug-in rule for the bandwidth $h$, which is similar to the famous rule of [17, Sec. 3.4]. Secondly, we introduce the readers to our $R$ package `logKDE`, which implements log-KDE in a manner that is familiar to users of the base $R$ function `density`. Thirdly, we provide details regarding a set of simulation studies and demonstrate the use of the log-KDE methodology via a pair of example data sets.

The paper proceeds as follows. Theoretical results for log-KDE are presented in Section 2. Use of the `logKDE` package is described in Section 3. Numerical studies and examples are detailed in Section 4.

## 2 Theoretical Results

We start by noting that MSE and MISE expressions for the log-KDE with lognormal kernels have been derived by [9]. The authors have also established the conditions for pointwise asymptotic unbiasedness and consistency for the lognormal kernel. In the general case, informal results regarding expressions for the pointwise bias and variance, have been provided by [4]. In this section, we generalize the results of [9] and formalize the results of [4] via some previously known results from [20] and [19, Sec. 2.5]. In the sequel, we shall make assumptions (A1) and (A2) regarding $f_Y(y)$, (A1*) and (A2) regarding $f_X(x)$, and (B1)–(B4) regarding $K(y)$.

### 2.1 Pointwise Results

The following expressions are taken from [19, Sec. 2.5]. At any $y \in \mathbb{R}$, define the pointwise bias and variance between (1) and $f_Y(y)$ as

$$\text{Bias}\left[\hat{f}(y)\right] = \mathbb{E}\left[\hat{f}(y)\right] - f_Y(y)$$
$$= \frac{1}{2}h^2 f_Y^{(2)}(y) + o(h^2), \qquad (3)$$

and

$$\text{Var}\left[\hat{f}(y)\right] = \frac{1}{nh} f_Y(y) \int_{\mathbb{R}} K^2(z)\, dz + o\left(\frac{1}{nh}\right), \qquad (4)$$

respectively, where $a_n = o(b_n)$ as $n \to \infty$, if and only if $\lim_{n\to\infty} |a_n/b_n| = 0$. From expressions (3) and (4), and the change-of-variable formula, we obtain the following expressions for the bias, variance, and MSE between (2) and $f_X(x)$.

**Proposition 1.** *For any $x \in (0, \infty)$, the bias, variance, and MSE between (2) and $f_X(x)$ have the forms*

$$Bias\left[\hat{f}_{\log}(x)\right] = \mathbb{E}\left[\hat{f}_{\log}(x)\right] - f_X(x) \tag{5}$$
$$= \frac{h^2}{2}\left[f_X(x) + 3xf_X^{(1)}(x) + x^2 f_X^{(2)}(x)\right] + o\left(h^2\right),$$

$$Var\left[\hat{f}_{\log}(x)\right] = \frac{1}{nhx} f_X(x) \int_{\mathbb{R}} K^2(z)\,dz + o\left(\frac{1}{nh}\right), \tag{6}$$

*and*

$$\begin{aligned}MSE\left[\hat{f}_{\log}(x)\right] &= Var\left[\hat{f}_{\log}(x)\right] + Bias^2\left[\hat{f}_{\log}(x)\right] \\ &= \frac{1}{nhx} f_X(x) \int_{\mathbb{R}} K^2(z)\,dz \\ &\quad + \frac{h^4}{4}\left[f_X(x) + 3xf_X^{(1)}(x) + x^2 f_X^{(2)}(x)\right]^2 \\ &\quad + o\left(\frac{1}{nh} + h^4\right),\end{aligned} \tag{7}$$

*respectively*

Consult cran.r-project.org/web/packages/logKDE/vignettes/logKDE.pdf for proofs of all theoretical results.

Let $h = h_n > 0$ be a positive sequence of bandwidths that satisfies the classical assumptions (D1) $\lim_{n\to\infty} h_n = 0$ and (D2) $\lim_{n\to\infty} nh_n = \infty$. That is, $h_n$ approaches zero at a rate that is slower than $n^{-1}$. Under (D1) and (D2), we have obtain the pointwise unbiasedness and consistency of (2) as an estimator for $f_X(x)$.

*Remark* 1. As noted by [4], the performance of the log-KDE method is most hindered by the behavior of the estimand $f_X(x)$, when $x = 0$, because of the $x^{-1} f_X(x)$ term in (6). If this expression is large at $x = 0$, then we can expect that the log-KDE will exhibit high levels of variability and a large number of observations $n$ may be required in order to mitigate such effects. From the bias expressions (5), we also observe influences from expressions of form $xf_X^{(1)}(x)$ and and $x^2 f_X^{(2)}(x)$. This implies that there may be a high amount of bias when estimating $f_X(x)$ at values where $x$ is large and $f_X(x)$ is either rapidly changing or the curvature of $f_X(x)$ is rapidly changing. Fortunately, in the majority of estimating problems over the domain $(0, \infty)$, both $f_X^{(1)}(x)$ and $f_X^{(2)}(x)$ tend to be decreasing in $x$, hence such effects should not be consequential.

### 2.2  Integrated Results

We denote the asymptotic MISE between a density estimator and an estimand as the AMISE. From the general results of [20], we have the identity

$$\text{MISE}\left[\hat{f}_{\log}\right] = \int_0^\infty \text{MSE}\left[\hat{f}_{\log}(x)\right] dx$$
$$= \text{AMISE}\left[\hat{f}_{\log}\right] + o\left(\frac{1}{nh} + h^4\right),$$

where

$$\text{AMISE}\left[\hat{f}_{\log}\right] = \frac{1}{nh}\mathbb{E}\left[X^{-1}\right]\int_{\mathbb{R}} K^2(z)\, dz$$
$$+ \frac{h^4}{4}\int_0^\infty \left[f_X(x) + 3xf_X^{(1)}(x) + x^2 f_X^{(2)}(x)\right]^2 dx.$$

By a standard argument

$$h^* = \arg\inf_{h>0}\ \text{AMISE}\left[\hat{f}_{\log}\right] \tag{8}$$
$$= \left[\frac{\mathbb{E}\left[X^{-1}\right]\int_{\mathbb{R}} K^2(z)\, dz}{\int_0^\infty \left[f_X(x) + 3xf_X^{(1)}(x) + x^2 f_X^{(2)}(x)\right]^2 dx}\right]^{1/5} n^{-1/5},$$

and

$$\inf_{h>0}\ \text{AMISE}\left[\hat{f}_{\log}\right] = \frac{5}{4}\left[\int_{\mathbb{R}} K^2(z)\, dz\right]^{4/5} J n^{-4/5}, \tag{9}$$

where

$$J = \left(\mathbb{E}^4\left[X^{-1}\right]\int_0^\infty \left[f_X(x) + 3xf_X^{(1)}(x) + x^2 f_X^{(2)}(x)\right]^2 dx\right)^{1/5}.$$

Using expression (8), we can derive a plugin bandwidth estimator for common interesting pairs of kernels $K(y)$ beand estimands $f_X(x)$. For example, we may particularly interesting in obtaining an optimal bandwidth $h^*$ for scenario where we take $K(y)$ to be normal and $f_X(x)$ to be log-normal with scale parameter $\sigma^2 > 0$ and location parameter $\mu \in \mathbb{R}$. This scenario is analogous to the famous rule of thumb from [17, Sec. 3.4].

**Proposition 2.** *Let $K(y)$ be normal, as per Table 1 and let $f_X(x)$ be log-normal, with the form*

$$f_X(x) = \frac{1}{x\sqrt{2\pi\sigma^2}}\exp\left(-\frac{1}{2\sigma^2}[\log x - \mu]^2\right).$$

If we estimate $f_X(x)$ by a log-KDE of form (2), then the bandwidth that minimizes $AMISE\left[\hat{f}_{\log}\right]$ is

$$h^* = \left[\frac{8\exp\left(\sigma^2/4\right)}{\sigma^4 + 4\sigma^2 + 12}\right]^{1/5} \frac{\sigma}{n^{1/5}}, \tag{10}$$

and

$$\inf_{h>0} \ AMISE\left[\hat{f}_{\log}\right] = \frac{5}{16}\left(\frac{2}{\pi}\right)^{2/5} n^{-4/5} J,$$

where

$$J = \frac{1}{2}\frac{\exp\left(9\sigma^2/20\right)\left(\sigma^4 + 4\sigma^2 + 12\right)^{1/5}}{\pi^{1/10}\sigma}.$$

*Remark 2.* In general, one does not know the true estimand $f_X(x)$, or else the problem of density estimation becomes trivialized. However, as a guideline, the log-normal density function can be taken as reasonably representative with respect to the class of densities over the $(0, \infty)$. As such, the plugin bandwidth estimator (10) can be used in order to obtain a log-KDE with reasonable AMISE value. Considering that the true parameter value $\sigma^2$ is also unknown, estimation of this quantity is also required before (10) can be made useful. If $\{X_i\}_{i=1}^n$ is a sample that arises from a log-normal density with parameters $\sigma^2$ and $\mu$, then $\{Y_i\}_{i=1}^n$ ($Y_i = \log X_i$, $i \in [n]$) is a sample that arises from a normal density with the same parameters. Thus, faced with $\{X_i\}_{i=1}^n$, one may take the logarithmic transformation of the data and compute the sample variance of the data to use as an estimate for $\sigma^2$. Alternatively, upon taking the logarithmic transformation, any estimator for $\sigma^2$ with good properties can be used. For example, one can use the interquartile range divided by $1.349^2$.

Rule (10) is by no means the only available technique for setting the bandwidth $h$ when performing log-KDE. An alternative to using rule (10) is to utilize the classic rule from [17, Sec. 3.4], based on minimizing the AMISE with respect to the estimator of form (1) using normal kernels, for estimating normal densities.

Apart from the two aforementioned plugin bandwidth estimators, we can also utilize more computationally intensive methodology for choosing the bandwidth $h$, such as cross-validation (CV) procedures that are discussed in [17, Ch. 3] or the improved efficiency estimator of [16]. The implementations of each of the mentioned methods for bandwidth selection in the `logKDE` package are discussed in further detail in the following section.

## 3 The `logKDE` package

The `logKDE` package can be installed from github and loaded into an active $R$ session using the following commands:

```
install.packages("logKDE", repos='http://cran.us.r-project.org')
```

The `logKDE` package seeks replicate the syntax and replicate the functionality of the KDE estimation function, `density`, built into the base $R$ `stats` package. The two main functions included in the package, `logdensity` and `logdensity_fft`, both return Density objects, which are of the same form as those produced by `density`. This enables the efficient reuse of functions such as `plot` and `print`, included in the `stats` package. Descriptions of all logKDE functions can be found in the manual [11].

### 3.1 Kernels

All of the kernels described in Table 1 are available in `logKDE`. They can be chosen via the `kernel` parameter in `logdensity` and `logdensity_fft`. The different options are `epanechnikov`, `gaussian`, `laplace`, `logistic`, `triangular`, and `uniform`. Note that `uniform` is referred to as `rectangular` in `stats`. By default we set `kernel='gaussian'` (i.e., log-normal). This choice was made to conform with the default settings of the `density` function.

### 3.2 Bandwidth selection

The available Bandwidth selection methods with the `logKDE` package include all of those from `stats` as well as two new bandwidth (BW) methods. The available methods are `bcv`, `bw.logCV`, `bw.logG`, `bw.SJ`, `nrd0`, `nrd`, and `ucv`. The equation of [17] is used to compute the `nrd0` bandwidth. The `nrd` bandwidth is the same as `nrd0`, except for a multiplicative constant. The bandwidths `ucv` and `bcv` are computed as per the descriptions of [15], performed on the log-transformed data. The `bw.SJ` bandwidth is computed as per the description of [16], and `bw.logG` bandwidth utilizes Equation (10). Finally, `bw.logCV` computes unbiased cross-validation bandwidths using untransformed data, rather than the log-transformed data that are used by `ucv`; see [4] for details.

### 3.3 Plotting and visualization

The reuse of functionality from base R packages allows intuitive visualization of the densities estimated using `logKDE`. This is illustrated in the following simple example (see Figure 1):

```
> fit1 <- logdensity(rchisq(100,10))
> plot(fit1)

> print(fit1)
Call:
logdensity(x = chisq10)
Data: chisq10 (100 obs.); Bandwidth 'bw' = 0.1546
       x                 y
 Min.   : 1.966   Min.   :0.003108
```

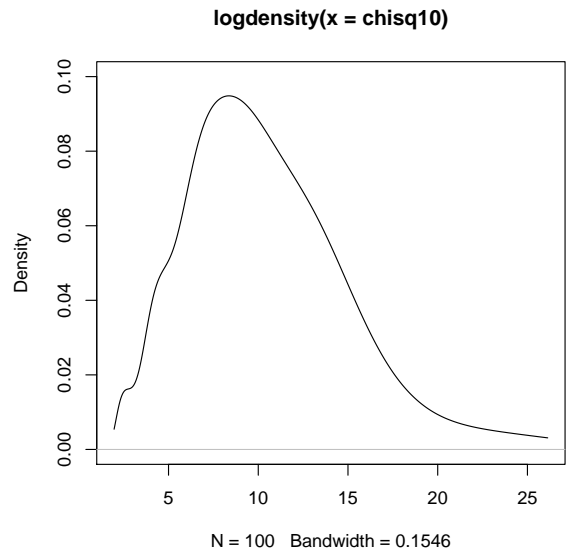

**Fig. 1.** A basic example of the use of `logdensity` class from R package `logKDE`.

```
 1st Qu.: 8.007    1st Qu.:0.008810
 Median :14.048    Median :0.034050
 Mean   :14.048    Mean   :0.040812
 3rd Qu.:20.089    3rd Qu.:0.071367
 Max.   :26.131    Max.   :0.094840
```

The shared syntax and class structure between `logdensity` and `density` allows for the simple creation of more complex graphical objects. Additionally, via a range of settings and options, different bandwidth and kernel preferences can be easily accessed (see Figure 2):

```
> fit<-logdensity(chisq10, bw ='logCV', kernel = 'triangular')
> plot(fit, ylim=c(0, 0.1))
> grid(10,10,2)
> x<-(seq(min(chisq10), max(chisq10), 0.01))
> lines(x, dchisq(x,10), col = 4)
```

## 4  Numerical Results

### 4.1  Simulation studies

A comprehensive set of simulation studies was conducted, based on the simulation studies of [4]. The performance of the The performance of the `logKDE` package was compared with those of the methods from `stats` and `Conake` [21]. The performances of the kernel density estimators from the various packages were compared via the average MISE and MIAE criteria. The results of the simulation studies that have been described are reported in Tables 2–7 of the `logKDE` vignette.

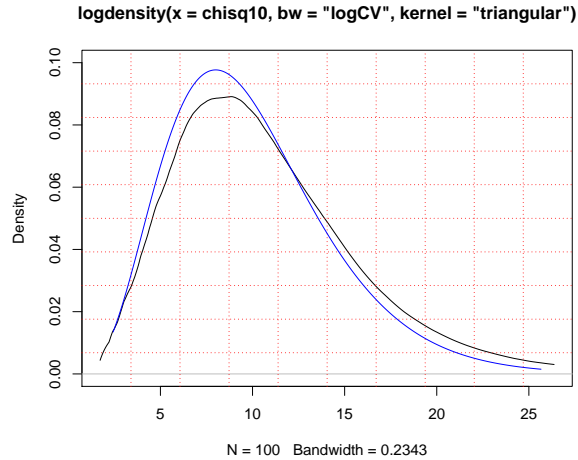

**Fig. 2.** Another example of the use of `logdensity` class from R package `logKDE`. In this case, the bandwidth is selected using the CV method and a triangular kernel is used. The $\chi^2_{10}$ reference distribution is marked in blue, whereas the log-KDE is plotted in black.

### 4.2   Case studies with real data

Our first illustrative example of the relative performance of the log-KDE method compared with standard KDE is provided using data taken from [23]. These data comprise of 331 salaries of Major League Baseball players for the 1992 season. Both densities were constructed using the default settings of the respective packages and both were estimated over the range [0.0001, 6500]. As can be seen in Figure 4, the log-KDE estimate is qualitatively closer to the histogram of the actual data, particularly for values that are close to the origin.

Another famous set of positive data is the daily ozone level data taken from a wider air-quality study [3]. The data consist of 116 daily measurements of ozone concentration in parts per billion taken in New York City, between May and September, 1973. The default settings and log-normal kernels were used for both estimators, which covered the range [0.0001, 200]. As with the baseball data, the fidelity of the kernel density estimate is improved, close to the origin.

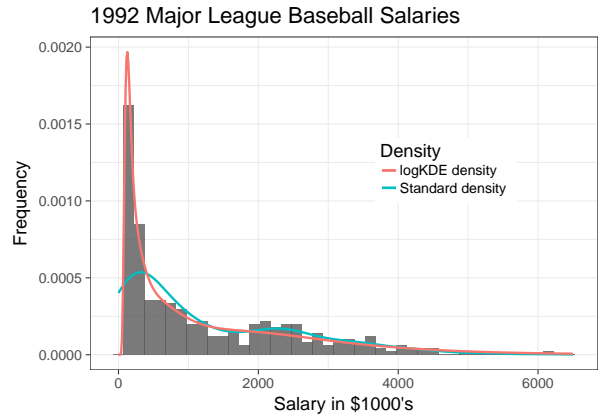

**Fig. 3.** Histogram and KDEs of the Baseball Salary data from [23].

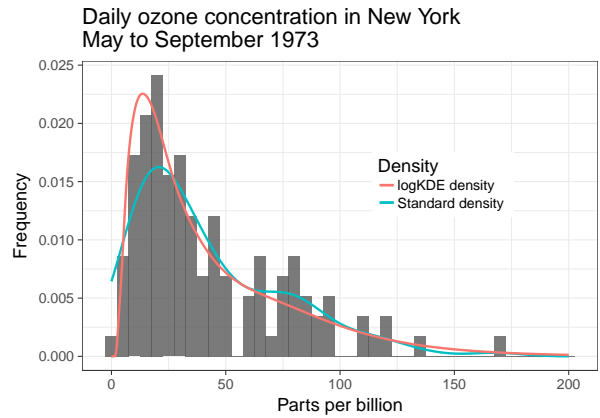

**Fig. 4.** Histogram and KDEs of the daily ozone concentration data taken from the air quality dataset in [3].

12      Andrew Jones et al.